# Quantitative Analysis of Light Induced Ion Segregation in Mixed-Halide Perovskites

**Petr Machovec,[a*] Lukáš Horák,[a] Milan Dopita,[a] Neda Neykova,[b,d] Lucie Landová,[b,d] Jakub Holovský,[b] Václav Holý,[a,c]**

[a] Faculty of Mathematics and Physics, Charles University, Ke Karlovu 3, Prague, Czech Republic

[b] Faculty for Electrical Engineering, Czech Technical University in Prague, Technická 2, 166 27 Prague, Czech Republic

[c] Faculty of science, Masaryk University, Kotlářská 2, 61137 Brno, Czech Republic

[d] Institute of Physics, Czech Academy of Sciences, v.v.i. Cukrovarnická 10, 162 00 Prague, Czech Republic

**Synopsis** A quantitative x-ray diffraction approach is introduced to resolve light-induced halide segregation in mixed-halide perovskite thin films, revealing the formation of Br-rich regions and their slow, incomplete relaxation in darkness.

**Abstract** Mixed-halide perovskites (MHPs) offer good band gap tunability by stoichiometry changes, which is an essential property for the creation of multijunction solar cells. However, under illumination, halide ions in MHP segregate and create I- and Br-rich regions, which decreases the efficiency of potential solar cells. In this work, a method for a detailed investigation of the distribution of halide ions within the MHP during and after illumination is introduced. Calculations of the strain field created by the halide segregation were performed, and the obtained local displacement of atoms was used to calculate the x-ray diffuse scattering. By fitting the experimental data measured on the thin polycrystalline layer of $FA_{0.83}Cs_{0.17}Pb(I_{0.6}Br_{0.4})_3$ the distribution of $Br^-$ and $I^-$ ions within an illuminated MHP was determined and the subsequent relaxation process of the segregation in the dark was tracked. Creation of highly Br-rich regions within slightly I-rich volume during the illumination was observed.

## 1. Introduction

Metal halide perovskites have attracted significant attention in recent years due to their exceptional properties, such as long carrier lifetime (Dequilettes *et al.*, 2016; Ahmed *et al.*, 2018) and high photoluminescence quantum yields (Sutter-Fella *et al.*, 2016; Ahmed *et al.*, 2018; De Roo *et al.*, 2016; Kroupa *et al.*, 2018). These optoelectronic properties combined with easy fabrication (Chen *et al.*, 2014; Jeon *et al.*, 2014; Mitzi *et al.*, 1963; Xiao *et al.*, 2014; Saidaminov *et al.*, 2015; Burschka *et al.*, 2013; Howard *et al.*, 2019) make metal halide perovskites promising candidates for applications in photovoltaics, LEDs, and photodetectors.

The ideal halide perovskite structure is cubic ABX$_3$, where A is a cation (MA+, FA+, Cs+), B is a metal cation (Pb2 +, Sn2 +, etc.), and X is a halide anion (Cl-, Br-, I-). However, the ideal cubic structure can readily be distorted into a tetragonal or orthorhombic phase, with the transition driven by ionic size difference (Johnsson & Lemmens, 2008; Pradeep *et al.*, 2024).

The flexibility of the perovskite structure allows for the mixing of different ions at similar lattice sites. Such modifications offer relatively easy access to band gap tunability, enhanced crystallinity, and improved chemical stability. The wide band-gap range, extending from approximately 1 eV for lead iodide perovskites (Sa *et al.*, 2020; Noh *et al.*, 2013) to over 2 eV for lead bromide perovskites (Datta *et al.*, 2025; Noh *et al.*, 2013; Eperon *et al.*, 2014; Amalathas *et al.*, 2025) and approaching 3 eV for lead chloride perovskites (Protesescu *et al.*, 2015; Lu *et al.*, 2025) makes perovskites ideal for both perovskite–perovskite and perovskite–silicon tandem photovoltaics.

However, the long-term stability of mixed halide perovskites (MHPs) under illumination remains a critical challenge that limits practical deployment. One of the primary instability mechanisms is photoinduced halide ion segregation, where light exposure causes halide ions to migrate and form regions with different stoichiometries, leading to phase separation (Hoke *et al.*, 2015; Barker *et al.*, 2017). This results in a redshift of the band gap, altering the optical properties, and reducing device efficiency over time. Before technologies based on MHPs can be commercially deployed, this instability needs to be addressed.

Numerous studies in recent years have tried to explain this instability. Models of structural rearrangement (Gottesman *et al.*, 2015), redistribution (DeQuilettes *et al.*, 2016), decomposition (Juarez-Perez *et al.*, 2018; Li *et al.*, 2017; Tang *et al.*, 2016), grain boundaries (Ridzoňová *et al.*, 2022), trap states (Motti *et al.*, 2019), polaron (Bischak *et al.*, 2017, 2018), and band-gap (Pavlovetc *et al.*, 2021) have been proposed. (Kerner *et al.*, 2021) Propose a unifying model by identifying halide oxidation as the event initiating the demixing. But there is still disagreement across many experimental and theoretical studies on the mechanism and reversibility of segregation and subsequent remixing in MHPs. Furthermore, contradictory results have been reported on the timescale of the changes. While some studies report fast segregation and relaxation(Motti *et al.*, 2019; Nie *et al.*, 2016) with a characteristic time of less than a minute, other studies report characteristic times in tens of minutes or hours(Kim *et al.*, 2018; Juarez-Perez *et al.*, 2018; Nie *et al.*, 2016; Gottesman *et al.*, 2015; Ridzoňová *et al.*, 2022).

In this work, we propose a method for the calculation of influence of halide segregation in MHPs on the x-ray diffraction (XRD) peak profile. XRD as a tool for the investigation of halide segregation has been used in many studies (Barker *et al.*, 2017; Hoke *et al.*, 2015; Sutter-Fella *et al.*, 2018; Knight *et al.*, 2021; Suchan *et al.*, 2023). But to the best of our knowledge, no one has tried to quantitatively analyse the XRD patterns to determine the distribution of the halide concentration in samples after illumination.

## 2. Computational Method

For the simulation of deformation in a single crystal or in an individual grain of polycrystal of a MHP, we assume that the deformation is caused solely by a random spatial fluctuation of the Br/I ratio, which can be characterised by local Br concentration $c_{Br}(\mathbf{r})$. The deformation is purely elastic, and in the simulation, we neglect the relaxation of internal stresses at the sample surface as well as plastic relaxation on grain boundaries, dislocations, and other defects. For the XRD calculations, it is convenient to introduce a local displacement field $\mathbf{u}(\mathbf{r})$, defined with respect to an ideal lattice corresponding to a homogeneous mean Br concentration. The procedure for calculating this displacement field is described in Appendix A. Within the context of this work, the term strain refers strictly to the deformation of this idealized homogeneous lattice. It should not be confused with the stress-free strain, although the two quantities are interconvertible when the local Br concentration is known.

For the description of x-ray diffuse scattering from a single crystal of finite size, we use a standard model of a mosaic crystal, in which we assume that the crystal consists of randomly placed and randomly rotated mosaic blocks (Ullrich *et al.*, 2004). For simplicity, we assume that the mosaic blocks are spherical with random radii $R$; the radii obey the Gamma distribution with the mean value $\langle R \rangle \equiv R_0$ and root mean square (rms) deviation $\sigma_R = R_0/\sqrt{m}$, where $m$ is the order of the Gamma distribution. The probability of finding two points $\mathbf{r}, \mathbf{r}'$ in the same block is

$$P(|\mathbf{r}-\mathbf{r}'|) = \int_0^\infty dR\, w_m(R, R_0) P_0(|\mathbf{r}-\mathbf{r}'|, R) \tag{1}$$

where $w_m(R, R_0) = m^m R^{m-1} \exp(-mR/R_0)/(\Gamma(m) R_0^m)$ is the probability density of the Gamma distribution with the order $m$, and for $m = 0$ we can express P as (Holý *et al.*, 1993)

$$P_0(|\mathbf{r}-\mathbf{r}'|, R) = \begin{cases} 1 - 3|\mathbf{r}-\mathbf{r}'|/(4R) + |\mathbf{r}-\mathbf{r}'|^3/(16R^3) & \text{for } |\mathbf{r}-\mathbf{r}'| \leq 2R \\ 0 & \text{for } |\mathbf{r}-\mathbf{r}'| > 2R \end{cases} \tag{2}$$

The random orientation of the blocks will be accounted for later. Since a powder-like curve is obtained from the calculated intensity by averaging over all orientations of the crystal lattice, the crystallographic orientation of the blocks has no influence on the final powder-like profile. Therefore, in our case the influence of mosaicity on the diffraction profile is only the broadening due to a finite size of the mosaic blocks.

Calculating diffuse x-ray scattering from a MHP crystal, we assume that (i) the kinematical approximation is valid, i.e., the blocks in the sample are much smaller than the x-ray extinction length (several microns), and (ii) the measured signal is averaged over all possible microstates (all block sizes and random functions of the Br concentration distribution $c_{Br}(r)$). The latter assumption is valid if the irradiated sample volume is much larger than $R$ and $\xi$; $\xi$ is the correlation length of the random function $c_{Br}(r)$.

The kinematical formula for the reciprocal-space distribution of the scattered intensity around the reciprocal-lattice point $\boldsymbol{h}$ reads (Ullrich *et al.*, 2004)

$$I_{\boldsymbol{h}}(\boldsymbol{q}) = A \int d^3\boldsymbol{r} \int d^3\boldsymbol{r}' P(\boldsymbol{r}-\boldsymbol{r}') \langle S_{\boldsymbol{h}}(c_{Br}(\boldsymbol{r})) S_{\boldsymbol{h}}^*(c_{Br}(\boldsymbol{r}')) e^{-i\boldsymbol{h}\cdot(\boldsymbol{u}(\boldsymbol{r})-\boldsymbol{u}(\boldsymbol{r}'))} \rangle e^{-i\boldsymbol{q}\cdot(\boldsymbol{r}-\boldsymbol{r}')} \quad (3)$$

where $A$ is a constant containing the intensity of primary beam, the polarization factor, and the Lorentz factor, among others, $\boldsymbol{q} \equiv \Delta \boldsymbol{Q} = \boldsymbol{Q} - \boldsymbol{h}$, where $\boldsymbol{Q} = \boldsymbol{K}_f - \boldsymbol{K}_i$ is the diffraction vector, $\boldsymbol{K}_i, \boldsymbol{K}_f$ are the wave vectors of the primary and scattered beams, and $S_{\boldsymbol{h}}$ is the structure factor of the MHP unit cell depending on the local Br concentration $c_{Br}(r)$:

$$S_{\boldsymbol{h}}(c_{Br}(\boldsymbol{r})) = S_{\boldsymbol{h}0}[1 - c_{Br}(\boldsymbol{r})] + S_{\boldsymbol{h}1} c_{Br}(\boldsymbol{r}) \quad (4)$$

$S_{\boldsymbol{h}0}, S_{\boldsymbol{h}1}$ are the structure factors of pure FA$_{0.83}$Cs$_{0.17}$PbI$_3$ and FA$_{0.83}$Cs$_{0.17}$PbBr$_3$ perovskites, respectively. The averaging $\langle \rangle$ runs over all random functions $c_{Br}(r)$, the reciprocal-lattice vector $\boldsymbol{h}$ is defined with respect to the MHP lattice with the mean Br concentration $\langle c_{Br} \rangle$.

From the practical point of view, it is convenient to split the argument of the integral in (3) to the diffuse and "coherent" parts

$$E_{\boldsymbol{h}}(\boldsymbol{r}) = S_{\boldsymbol{h}}(\boldsymbol{r}) e^{-i\boldsymbol{h}\cdot\boldsymbol{u}(\boldsymbol{r})} = [S_{\boldsymbol{h}}(\boldsymbol{r}) e^{-i\boldsymbol{h}\cdot\boldsymbol{u}(\boldsymbol{r})} - \langle S_{\boldsymbol{h}} \rangle] + \langle S_{\boldsymbol{h}} \rangle \equiv E_{\boldsymbol{h}}^{(diff)}(\boldsymbol{r}) + E_{\boldsymbol{h}}^{(coh)} \quad (5)$$

Then, after some manipulation, we obtain the following final formula, which is suitable for numerical simulations.

$$I_h(\boldsymbol{q}) = A\left[\frac{1}{8\pi^3}P^{(FT)}(\boldsymbol{q}) \otimes \langle|E_h^{(diff,FT)}(\boldsymbol{q})|^2\rangle + VP^{(FT)}(\boldsymbol{q})|E_h^{(coh)}|^2 \right.$$
$$\left. + 2Re\left(P^{(FT)}(\boldsymbol{q})\langle E_h^{(diff,FT)}(0)\rangle E_h^{(coh)*}\right)\right], \tag{6}$$

where $V$ is the irradiated sample volume and $\otimes$ means convolution, and $^{FT}$ stands for Fourier transformation.

In each simulation, we generate N realizations of the random function $c_{Br}(\boldsymbol{r})$ assuming the correlation function in the form

$$\langle(c_{Br}(\boldsymbol{r}) - \langle c_{Br}\rangle)(c_{Br}(\boldsymbol{r}') - \langle c_{Br}\rangle)\rangle = \sigma^2 \exp\left(-\frac{|\boldsymbol{r}-\boldsymbol{r}'|^2}{\xi^2}\right), \tag{7}$$

with root-mean square (rms) deviation $\sigma = \sqrt{\langle(c_{Br} - \langle c_{Br}\rangle)^2\rangle}$ and given correlation length $\xi$. Additionally, we introduce an asymmetry of the random fluctuation. This enables us to modify the distribution of concentration. Specifically, it facilitates the creation of a concentration distribution characterized by small volumes of elevated concentration alongside larger volumes where the concentration is only marginally lower than the mean and vice versa. This asymmetry represents a situation in which bromine or iodine ions from a larger area are attracted to specific centres.

In the first step of a simulation run the random fluctuation $c_{Br}(\boldsymbol{r})$ described by equation (7) is generated. Utilizing this concentration fluctuation $c_{Br}(\boldsymbol{r})$, a modified concentration $c_{Br}^{\alpha}(\boldsymbol{r})$ is constructed as follows:

$$c_{Br}^{\alpha}(\boldsymbol{r}) = \begin{array}{l}(c_{Br}(\boldsymbol{r}) - \langle c_{Br}\rangle) \text{ for all } \boldsymbol{r}, \text{where } c_{Br}(\boldsymbol{r}) \leq \langle c_{Br}\rangle \\ (c_{Br}(\boldsymbol{r}) - \langle c_{Br}\rangle)\cdot\alpha \text{ for all } \boldsymbol{r}, \text{where } c_{Br}(\boldsymbol{r}) > \langle c_{Br}\rangle\end{array} \tag{8}$$

where the parameter $\alpha$ from interval $(0, \infty)$ is the parameter of asymmetry. The final asymmetrical fluctuation in bromine concentration, denoted as $c_{Br}^{asym}(\boldsymbol{r})$ is then expressed as:

$$c_{Br}^{asym}(\boldsymbol{r}) = (c_{Br}^{\alpha}(\boldsymbol{r}) - \langle c_{Br}^{\alpha}(\boldsymbol{r})\rangle)\cdot\frac{\sigma}{\Sigma} + \langle c_{Br}\rangle \tag{9}$$

where $\sigma$ is the original desired standard deviation and $\Sigma$ is the standard deviation of $(c_{Br}^{\alpha}(\boldsymbol{r}) - \langle c_{Br}^{\alpha}(\boldsymbol{r})\rangle)$. This ensures preservation of correct mean value and standard deviation of the fluctuation. The parameter α acts as an amplitude-scaling factor for the random fluctuations. Values $\alpha > 1$ enhance the positive deviations and suppress the negative ones, leading to higher maxima and shallower minima. Conversely, values $\alpha < 1$ attenuate the positive deviations and deepen the negative ones, resulting in lower maxima and deeper minima. To elucidate the role of $\alpha$, Figure 1 presents representative examples of the resulting distributions together with their cumulative histograms.

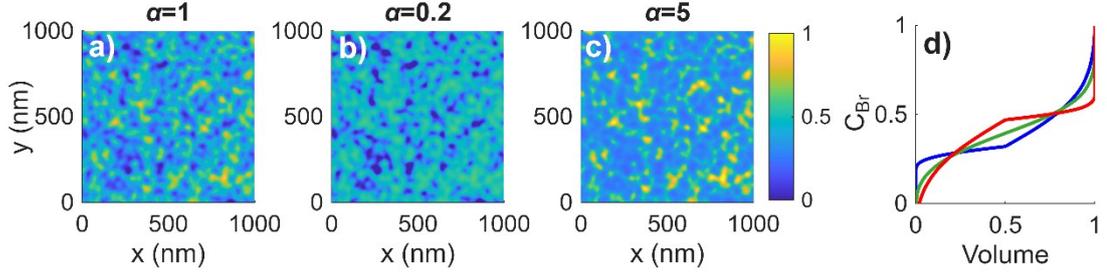

**Figure 1** Examples of two dimensional concentration profiles of Br generated with the correlation function in Eq. (7) for a) $\alpha = 1$, b) $\alpha = 0.2$, and c) $\alpha = 5$ with correlation length $\xi = 30$ nm and root mean square deviation $\sigma = 0.15$. The mean Br concentration for all distributions is 0.4. In panel d) we present normalised cumulative histogram of concentration within the sample, that expresses in what part of the volume is at most the given concentration of Br for $\alpha = 1$ (green), $\alpha = 5$ (blue), and $\alpha = 0.2$ (red).

We calculate the final mean intensity as an average of the intensities calculated for different realizations of $c_{Br}(\mathbf{r})$.

$$I_{\mathbf{h}}(\mathbf{q}) = \frac{1}{N}\sum_{i=1}^{N} I_{\mathbf{h}}^{(i)}(\mathbf{q}). \qquad (10)$$

The number of configurations to average over, denoted as $N$, must be selected such that the difference in the final intensity between $N$ and $N + 1$ is negligible. It has been determined that $N = 100$ meets this criterion while concurrently being small enough to facilitate reasonable calculation times and therefore it was used for all our calculations.

To obtain powder-like curve from the three-dimensional distribution of the intensity, we numerically integrate each function $I_{\mathbf{h}}^{(i)}(\mathbf{q})$ along the direction perpendicular to $\mathbf{h}$, approximating the integration over a Debye sphere, and calculate the final intensity by (10).

For the following considerations it is useful to express $E_{\mathbf{h}}^{(diff)}(\mathbf{r})$ defined in equation (5) as

$$E_{\mathbf{h}}^{(diff)}(\mathbf{r}) = S_{\mathbf{h}}(\mathbf{r})e^{-i\mathbf{h}.\mathbf{u}(\mathbf{r})} - \langle S_{\mathbf{h}}\rangle = \langle S_{\mathbf{h}}\rangle e^{-i\mathbf{h}.\mathbf{u}(\mathbf{r})} + \Delta S_{\mathbf{h}}(\mathbf{r})e^{-i\mathbf{h}.\mathbf{u}(\mathbf{r})} - \langle S_{\mathbf{h}}\rangle \qquad (11)$$

Where $\Delta S_{\mathbf{h}}(\mathbf{r})$ is local deviation of the structure factor due to the local deviation in Br concentration. The two distinct contributions to $E_{\mathbf{h}}^{(diff)}(\mathbf{r})$ are the chemical contrast $\Delta S_{\mathbf{h}}(\mathbf{r})e^{-i\mathbf{h}.\mathbf{u}(\mathbf{r})}$ and contribution from fluctuating strain $\langle S_{\mathbf{h}}\rangle(e^{-i\mathbf{h}.\mathbf{u}(\mathbf{r})} - 1)$. Here we separate these contributions only as a conceptual decomposition to explain the origin of the peak shape, whereas experimentally the measured intensity contains both inseparably. To illustrate the influence of the separate and combined contributions of chemical contrast, strain, both effects simultaneously and mosaicity we show corresponding reciprocal space maps (RSMs) simulated for random distribution of concentration with parameters $\sigma = 0.05$, $\alpha = 1$, and $\xi = 1$ nm in Figure 2.

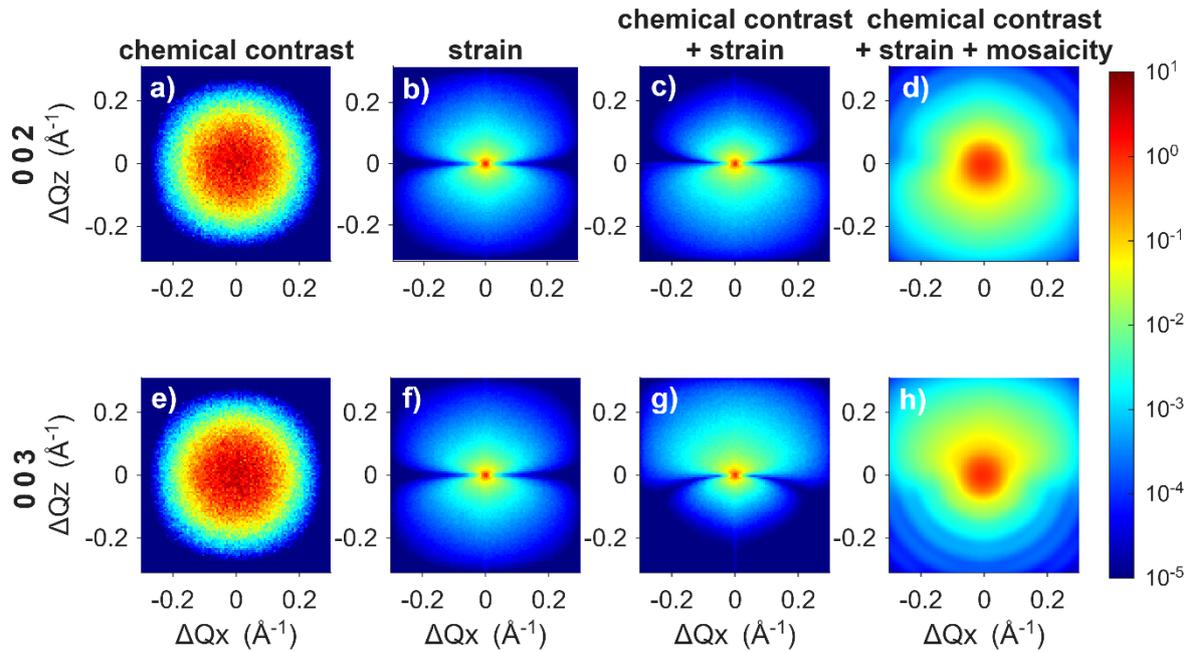

**Figure 2** Simulated reciprocal space maps of 0 0 2 a)-d) and 0 0 3 e)-h) diffraction maxima with a) and e) only chemical contrast included, b) and f) only strain included, c) and g) both chemical contrast and strain included, and d) and h) chemical contrast, strain and mosaicity included. The parameters of the random distribution of concentration are $\sigma = 0.05$, $\alpha = 1$, and $\xi = 1$ nm, and mosaic block size in d) and h) is $R = 5$ nm.

While both chemical contrast and strain separately yield centrosymmetric intensity distributions, when both chemical contrast and strain are included, the resulting RSMs exhibits strong asymmetry. To understand the origin of this asymmetry, we will consider a simple distribution of $c_{Br}(\mathbf{r})$, specifically a spherical inclusion of either high or low Br concentration with a gaussian profile of concentration. The parameters of this inclusion are maximal deviation of concentration of 0.1 from the average concentration $c_{Br} = 0.4$ and standard deviation of the gaussian distribution 10 nm. As shown in Figure 3 the absolute values of the strain term create two symmetrical lobes, but the phases of these lobes differ by $\pi$. Using equations (11) and (6) the symmetricity of the lobes can be demonstrated for both, the distribution of concentration with a spherical inclusion as well as for a distribution of concentration with correlation function described by equation (7). Full explanation is given in Appendix B. Therefore, one lobe has opposite phase than the other lobe and the signs of the phases depend on whether the inclusion has high Br (tensile strain within the inclusion) or I (compressive strain within the inclusion) concentration. Absolute values of term $\Delta S_h(r)e^{-ih.u(r)}$ create symmetrical spherical distribution but the phases of these spheres differ approximately by $\pi$ for different diffraction maxima. As a result, when these two terms are added in one lobe the absolute values are added and in the other subtracted, leading to an asymmetrical distribution of intensity. Therefore, the direction of this asymmetry is governed by the phase of the deviation of the structure factor $\Delta S_h$, which depends on the diffraction indices. A simple rule can be derived: for reflections with all Miller indices odd or all even, the asymmetry appears toward lower $Q$, whereas for mixed indices, it shifts toward higher $Q$. This explanation is in many ways analogous to Huang scattering (Krivoglaz, 1996), where a contribution of the diffuse scattering from the deformed area around a defect core and the contribution of the defect core itself are superposed. In Huang scattering the asymmetry has the same sign for all diffraction maxima and is determined simply by the sign of the strain.

Both high and low Br (low and high I) concentration inclusions create the asymmetry in the same direction because the phases of both the strain and the chemical contrast flip signs for high versus low Br concentration. Therefore, as shown in Figure 2, even for a random fluctuation, that essentially consist of high and low Br concentration inclusions, we observe the described asymmetry.

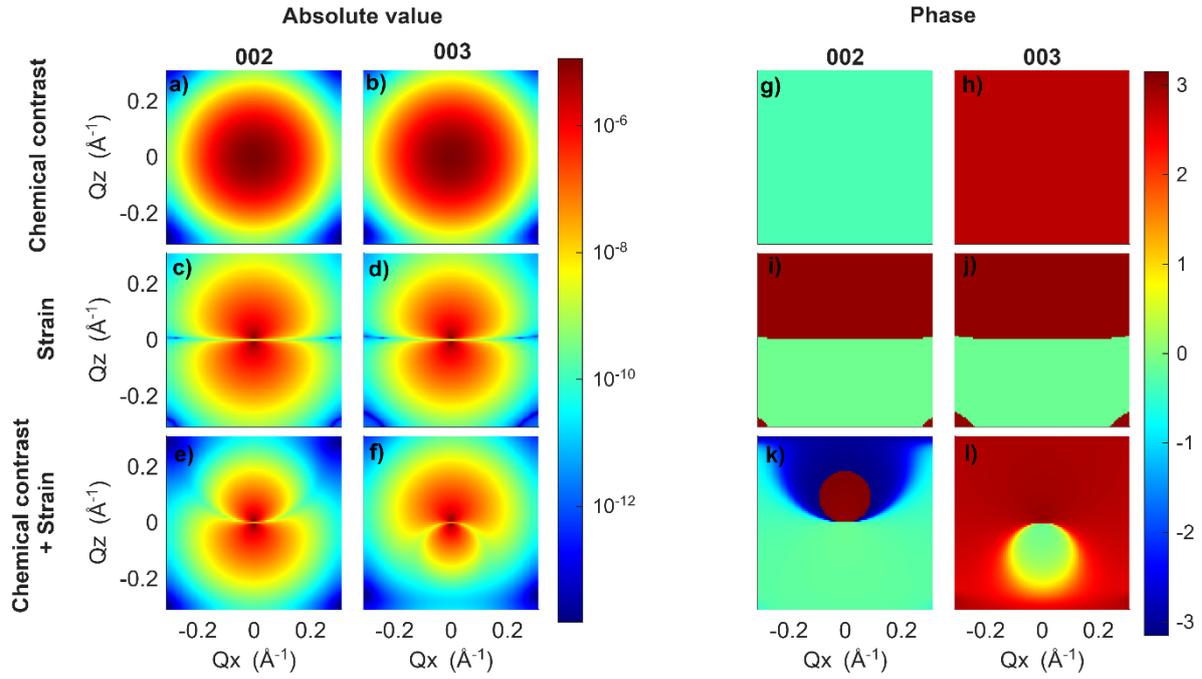

**Figure 3** Absolute values a)-f) and phases g)-l) of simulated structure factors of spherical inclusion of high Br concentration with Gaussian profile. Pictures a), b), g), and h) show the structure factor considering only chemical contrast, pictures c), d), i), and j) show the structure factor considering only strain, and e), f), k), and l) show the structure factor considering both chemical contrast and strain.

To demonstrate the effects of different parameters on the simulated powder-like diffraction profiles and to evaluate the sensitivity of the model to individual parameters, a series of simulations was performed for various parameter values. The influence of different parameters is illustrated in Figure 4. As expected with increasing $\sigma$ we observe broadening of the profiles accompanied by the decrease of the maximum intensity. The asymmetry of diffraction maxima described in previous paragraph is visible for the symmetrical random fluctuation of the concentration ($\alpha = 1$).

For the asymmetrical random fluctuation of the concentration ($\alpha \neq 1$) the $\Delta S_h$ asymmetry is mostly covered by the asymmetry of the concentration. This is demonstrated in Figure 4e)-h), for $\alpha \neq 1$ we observe asymmetry of all diffraction maxima to the same side. For $\alpha < 1$ the asymmetry is to lower $Q$ (not shown here) and for $\alpha > 1$ the asymmetry is to higher $Q$. The model is sensitive to the values of $\alpha$ from the interval $\langle 0.2, 5 \rangle$, outside of this interval the peak shape no longer changes with changing $\alpha$. This asymmetry is accompanied by a decrease in maximum intensity and a slight shift in the position of the maximum intensity. This shift is understandable, because most of the volume now has slightly higher or lower Br concentration and therefore slightly lower or higher lattice parameters.

The dependence of the profile shape on the correlation length is demonstrated in Figure 4i)-l). For correlation length under 7 nm a sharp peak is present, for longer correlation length the profiles are broader, and the central sharp peak disappears. Furthermore, it is important to note that for the correlation length above 15 nm the shape of the peak is no longer influenced by changes in $\xi$.

As it follows from our analysis, no particular hkl reflections are intrinsically more sensitive to specific parameters. Nevertheless, the dependence of peak broadening on the scattering-vector magnitude differs among parameters. Consequently, measuring and fitting as many reflections as possible is essential to enhance sensitivity and reduce parameter correlations. For example, reliable determination of $\alpha$ requires reflections with both equal-parity and mixed-parity Miller indices.

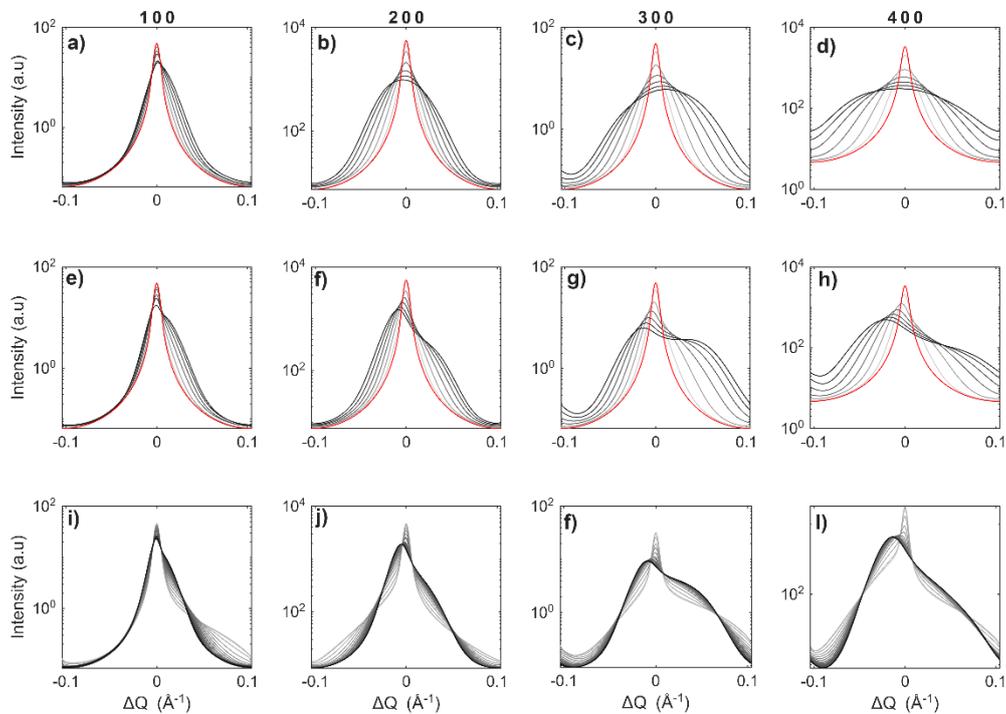

**Figure 4** Influence of root mean square deviation of the concentration on powder-like diffraction profiles a)-h), the red curves are profiles with uniform concentration. For the grey curves values of $\sigma$

go from 0.05 (light grey) to 0.3 (black). In plots a)-d) the random fluctuation of the concentration is symmetrical ($\alpha = 1$) and in e)-h) the random fluctuation of the concentration prioritises small regions with much higher Br concentration than the average concentration and large regions with slightly lower Br concentration than the average concentration ($\alpha = 5$). The plots i)-l) show the influence of correlation length on the diffraction profile, the values go from $\xi = 0.5$ nm (light grey) to 17.5 nm (black) in 0.5 nm increments. All shown simulations are for mosaic block size 50 nm.

### 3. Experimental

### 3.1. Sample Preparation

To demonstrate this method, thin films with chemical composition $FA_{0.83}Cs_{0.17}Pb(I_{0.6}Br_{0.4})_3$ were prepared on glass and silicon substrates using a one-step antisolvent approach. The precursor solution was obtained by dissolving the cation sources (FAI and CsI), along with $PbI_2$ (X mmol) and $PbBr_2$ (1-X mmol), in 1 mL of a DMF:DMSO (4:1) solvent mixture, followed by continuous stirring at 70 °C overnight.

The spin-coating process consisted of two sequential steps at different rotation speeds – typically 1000 rpm for 10 seconds, followed by 5000 rpm for 30 seconds. To influence the surface morphology and microstructure, an antisolvent (chlorobenzene, typically 150 µL) was dropped onto the spinning substrate 15 seconds before the end of the second step. After deposition, the films were annealed at 100 °C, then placed on an aluminum-foil-covered Petri dish and allowed to cool to room temperature for a few minutes. All fabrication steps were conducted inside a nitrogen-filled glovebox to prevent exposure to moisture and oxygen. The sample preparation is more closely described in (Ridzoňová *et al.*, 2022; Horynova *et al.*, 2025; Amalathas *et al.*, 2025)

### 3.2. X-ray diffraction

XRD measurements were performed using a Rigaku Smartlab diffractometer equipped with a rotating copper anode in parallel beam geometry with fixed incidence angle 1°. The angular resolution 0.5° in diffraction plane was defined by a parallel-plate analyser in front of the detector. The axial divergence and acceptance were limited to 5° by the Soller slits on the primary and the secondary side, respectively. The parallel beam geometry is suitable for this measurement because the low angle of incidence enhances the signal from the layer, compared to the substrate and the fixed angle of incidence ensures irradiation of the full surface area of the sample. The sample was kept in a PEEK dome in an argon atmosphere to eliminate potential decomposition of the sample caused by exposure to air and humidity. Outside of the well-defined illumination intervals, the sample was kept in the dark all time. For the measurement, we chose a $2\theta$ interval that contains diffraction peaks 1 0 0, 1 1 0, 1 1 1, 2 0 0, 2 1 0, 2 1 1, 2 2 0, 2 2 1. We chose a measurement time for each point to be 0.6 s, which, in combination with scanning line detector mode, results in each measurement taking 70 minutes. This gives us a reasonable quality of the data and at the same time, gives us a good enough time resolution to observe the relaxation of segregation in our sample. The measurements were conducted in a sequential manner, whereby each subsequent scan started immediately upon the completion of the preceding scan.

### 3.3. Sample Illumination

Illumination of the sample was done by a solar simulator PhotoFluor II by 89 North. Power of the solar simulator and distance to the sample were set such that the illumination of the sample was equivalent to an illumination by 1 Sun as calibrated by photocurrent on a silicon detector. During the illumination we did not perform any diffraction measurements.

During illumination, the sample is heated by irradiation; however, it rapidly cools back to room temperature in the dark. Therefore, we expect to see no diffraction-peak shift due to the thermal expansion.

### 4. Results

Before the first illumination, we performed a measurement of a pristine sample. This measurement confirmed a cubic perovskite structure with lattice parameter $a = 6.17$ Å and no macroscopic residual strain has been observed. Then we illuminated the sample for 10 minutes and immediately after illumination, we started a series of measurements to investigate the relaxation of photoinduced changes. After 53 hours of relaxation, we illuminated the sample once more, this time for 30 minutes, and measured the relaxation again.

For comparison of the simulated and measured data, we had to take into account several additional instrumental effects. First, the radiation passing through the x-ray mirror is not strictly monochromatic, but contains two characteristic wavelengths Cu K$\alpha_1$ $\lambda = 1.5406$ Å and Cu K$\alpha_2$ $\lambda = 1.5444$ Å. This is implemented by adding a scale copy of the peak to the appropriate 2θ position, according to the known wavelength separation and intensity ratio. Furthermore, the calculated diffraction curve was convoluted with the instrumental resolution function obtained from the measurement of LaB$_6$ standard. Finally, a linear background was added separately for each peak.

In the final model, the fitted parameters are the root mean square deviation of the concentration $\sigma$, the correlation length of the concentration $\xi$, the radius of the grains $R$, the factor of asymmetry $\alpha$, and a lattice parameter $a$. Additionally, for each maximum, a scale parameter was added to account for the difference in measured and predicted intensity of individual peaks. These scale parameters including for example the texture and geometrical influence of the diffractometer do not change for subsequent measurements. All peaks of the given diffraction pattern were fitted together with the same parameters.

In Figure 5 we present all diffraction patterns measured before and after illuminations. Before the illumination of the sample, we observe symmetrical diffraction maxima. After the short 10-minute illumination, the peak shifts slightly to lower $2\theta$ and decreases in intensity, but the most significant change is the new asymmetrical broadening of the peaks. All peaks broadened more in the direction of higher $2\theta$, and peaks at higher diffraction angles have a greater broadening. Over time, these changes relax, and after 48 hours, the peak shapes and positions are almost back to the original values. After this, the sample was illuminated once again, this time for 30 minutes. We observe similar behaviour as after the first illumination. During the entire procedure, we do not observe creation or annihilation of any diffraction maxima. In addition to the expected diffraction maxima of the phase $FA_{0.83}Cs_{0.17}Pb(I_{0.6}Br_{0.4})_3$ we also observe a peak at $2\theta = 12.7°$. A single peak is insufficient for definitive phase identification; however, its position is consistent with the (001) reflection of hexagonal $PbI_2$ (space group P-3m1) which is a common impurity phase observed in MHPs (Macpherson *et al.*, 2022). This peak does not change its intensity, position or shape during illumination/relaxation, indicating that this impurity remains stable under our conditions.

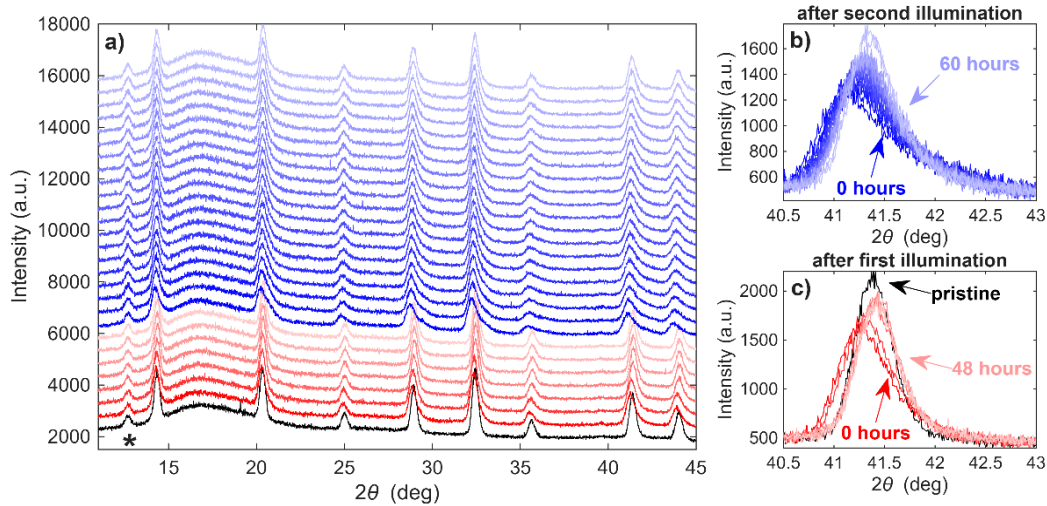

**Figure 5** a) Measured diffraction patterns before illumination (black), after 10 minutes of illumination (red), and another 30 minutes of illumination (blue), the data are vertically shifted for visibility. The panels b) and c) show details of diffraction maximum 2 2 0 after first and second illumination. The diffraction profiles of individual maxima after the illumination are broadened and asymmetrical, this effect gradually relaxes when the sample is kept in the dark. The peak denoted by an asterisk belongs to $PbI_2$ impurity.

We fitted all diffraction patterns by our model. In Figure 6 you can see a comparison of the measured data and the fit for a diffraction pattern measured after first illumination, for other fitted XRD patterns see Figure S1. For comparison of the simulation and the measured data, we recalculated the $2\theta$ angle to reciprocal space vector $Q$. Our model predicts well all described effects. Root mean square deviations of concentration $\sigma$ for all measurements, shown in Figure 7a), increases during illumination and exponentially relaxes in darkness. The mean radius of the crystalline grains is 50 nm for all measurements. The crystallite size is in good agreement with the values determined by scanning electron microscopy (Figure S2). The correlation length in all measurements is higher than the limit of sensitivity of the simulation which is 15 nm. We assume the correlation length to be the same as the crystallite size. And finally, for all measurements we observe positive asymmetry factor of value higher than 5, which is the sensitivity limit of the model to this parameter. This means that there are bromine-rich inclusions in a volume that have a slightly higher iodine concentration than the uniform concentration. This asymmetrical fluctuation of concentration is the only possible distribution of concentration that can explain the asymmetrical broadening of the diffraction maxima that is hkl independent and in the direction of higher $Q$. Within the presented model, an asymmetric concentration distribution ($\alpha > 1$) is required to reproduce an hkl-independent asymmetric broadening toward higher $Q$, whereas symmetric fluctuations ($\alpha = 1$) always produce an hkl-dependent asymmetry direction.

While majority of the observed shift of the maxima can be explained by our model (see Figure 4e-h)) we do observe small change of lattice parameter shown in Figure 7b), that can't be explained by our model. It also cannot be attributed to temperature changes as we expect that thermalization is finished before the measurement of the first peak. This is confirmed by the observation of no changes in the lattice parameters for different maxima during the initial measurement following light soaking.

For the explanation of this additional change of the lattice parameter we will consider a spherical highly Br-rich inclusion in slightly I-rich volume, which serves as a simplification of the distribution obtained by the fitting. The inclusion is expected to exhibit a lattice parameter smaller than that of the surrounding matrix. and given the requirement for the conservation of number of unit cell, the inclusion will try to contract compared to its original volume with the mean Br concentration, this will lead to strain in both the inclusion and the surrounding volume. This strain can be calculated by the theory developed by (Eshelby, 1957) and the mean lattice parameter of such a system is larger than the lattice parameter of the perovskite with mean Br concentration. Additionally, the lattice parameter may be further affected by the incorporation of Br or I ions at grain boundaries, within lattice defects, or in other regions that do not contribute to coherent diffraction. This changes the mean Br concentration within the coherently diffracting domains and therefore influences the mean lattice parameter.

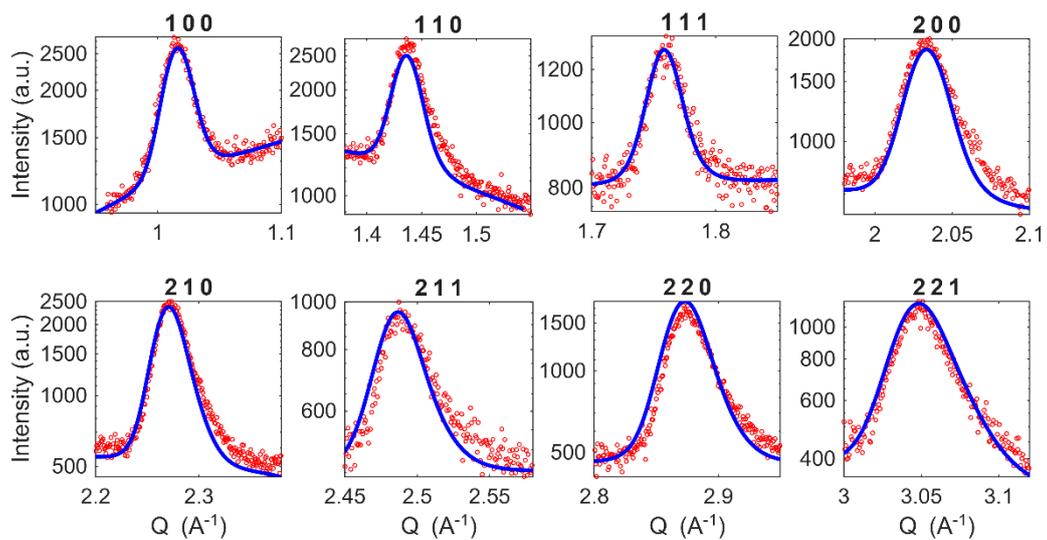

**Figure 6** Diffraction maxima measured after 10 minutes of illumination (red) and fits by our model (blue). The y-axis scale is logarithmic. The parameters obtained from the fitting are $\sigma = 0.13 \pm 0.01$, $\xi \geq 15\ nm$, $R = 50\ nm$, and $\alpha \geq 5$.

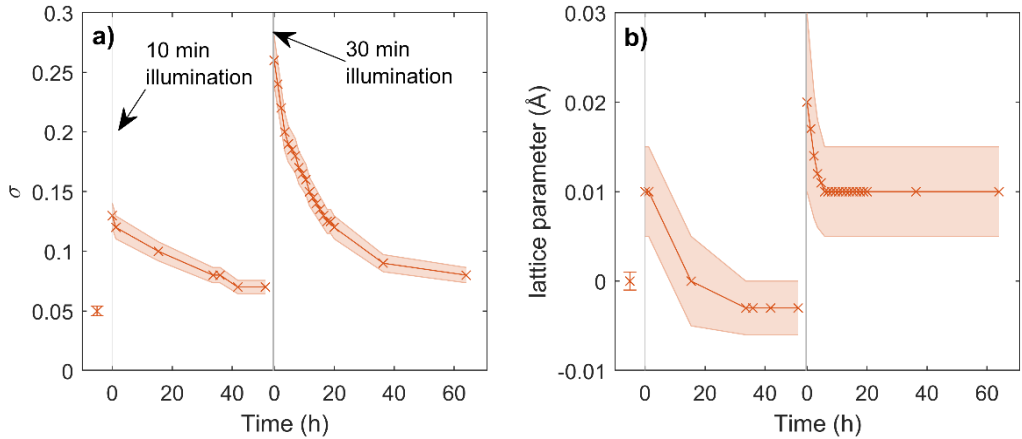

**Figure 7** Evolution of a) Root mean square deviation of concentration and b) lattice parameter during relaxation of segregation. The shaded areas represent error bars and grey areas mark the time of illumination of the sample. The x-axes show the time since the last illumination (relaxation time).

## 5. Discussion

Using the presented method, we were able to successfully determine the distribution of halide ions within the MHP during illumination and relaxation. This method is a complementary approach to photoluminescence spectroscopy (PL). PL probes primarily band gap of the volume with the lowest amount of defects in the MHP material and principally, due to thermodynamics, does not see the Br-rich region. XRD on the other hand investigates crystal structure and real structure in whole material volume equally.

In this work we have not considered the strain relaxation on surfaces, defects, and grain boundaries. Relaxation of the strain will lead to a smaller volume force density in Equation (A2) and therefore to smaller local displacement. This implies a limitation of our model, particularly in systems characterized by a high density of strain relaxation centres, such as grain boundaries. In such contexts, the model tends to underestimate the root mean square deviation of Br concentration. This can be accounted for by adding a variable that describes the relaxation into equation (A3) in the Appendix A. The question remains how to quantify the relaxation on different types of relaxation centres. Consequently, it was shown by theoretical calculations using Cahn-Hilliard equation that the strain relaxation is in fact needed for the halide demixing (Holý et al., 2025).

We observed creation of small Br-rich regions and larger slightly I-rich regions during illumination and subsequent slow and imperfect relaxation in the dark at the timescale of hours. Similar concentration distribution could also be represented by spherical inclusions of high concentration in a volume of slightly low concentration. However, to describe such an arrangement, we need much more parameters which makes the fitting cumbersome and ambiguous. Our efforts to describe the measured data by various alterations of spherical inclusions yielded significantly worse fits than the random concentration model.

Although most published studies agree on the timescale of the processes, not a single study explicitly reports the creation of highly $Br$-rich perovskite regions in a slightly $I$-rich volume. For the studies using only PL as a probe of local halide concentration, this is logical, as the $I$-rich regions have higher PL activity (due to a lower bandgap of $FAPbI_3$ (Tang et al., 2020) than that of $FAPbBr_3$ (Zhang et al., 2018)) and therefore dominate the spectra. By means of HRTEM (Funk et al., 2023) reveal creation of sub 5 nm Br-rich inclusion in $CsPb(Br_xI_{1-x})_3$. However, due to the structural similarity of $CsPbBr_3$ and $PbBr_2$, they are unable to decisively distinguish the phase and based on previously published research attribute these regions to the $PbBr_2$ phase. And in the few published papers using XRD to investigate halide segregation (Knight et al., 2021; Sutter-Fella et al., 2018) in $FA_yCs_{1-y}Pb(Br_xI_{1-x})_3$ we do not see any clear agreement on the halide distribution in illuminated samples. For the case of $MAPb(I_xBr_{1-x})_3$ perovskites, although not directly stated (Duong et al., 2017), observe splitting of the diffraction peaks that is consistent with creation of small Br-rich regions within slightly I-rich volume.

While the difference between $I^-$ and $Br^-$ migration activation energies in $FA_yCs_{1-y}Pb(Br_xI_{1-x})_3$ is smaller than in MA based perovskites(Oranskaia et al., 2018) the migration barrier for $I^-$ is still lower than $Br^-$. This means that observed creation of highly $Br$-rich regions in low $I$-rich volume suggests that the $I^-$ ions migrated outwards from some nucleation site such as grain boundaries, interfaces, or lattice defects.

## 6. Conclusions

In this work, we developed and validated a quantitative XRD method capable of resolving light induced halide segregation in mixed-halide perovskites. By combining elastic strain field calculations with a diffuse-scattering model, we were able to reconstruct the spatial distribution of Br and I ions within $FA_{0.83}Cs_{0.17}Pb(I_{0.6}Br_{0.4})_3$ thin films after illumination and subsequent relaxation in the dark. The model successfully reproduces illumination induced broadening, asymmetry, and shifts of diffraction maxima and enables the determination of the root mean square fluctuation, asymmetry, and correlation length of the halide concentration.

Our analysis reveals that illumination leads to the formation of highly Br-rich regions embedded within a slightly I-rich matrix, followed by a slow, incomplete relaxation over hours in darkness. These findings provide direct structural evidence for asymmetric halide redistribution. The presented approach offers a robust, complementary alternative to photoluminescence-based techniques, enabling bulk-sensitive and composition-specific insight into the mechanisms of light-induced demixing in metal halide perovskites.

**Acknowledgements**

**Conflicts of interest**    The authors declare that they have no conflict of interest.

**Data availability**    The data supporting the findings of this study are available from the corresponding author upon request.

**Funding information**    This work was supported by Czech Science Foundation project number GACR 23-06543S. and the Czech Ministry of Education, Youth and Sports grant no. CZ.02.01.01/00/22_008/0004617 – "Energy conversion and storage".

**Appendix A. Strain Field Calculation**

We assume that the MHP crystal is cubic and that its lattice parameter linearly depends on the local Br concentration $c_{Br}(\mathbf{r})$:

$$a(c_{Br}) = a_0(1 + \eta c_{Br}), \eta = \frac{a_1 - a_0}{a_0} \tag{A1}$$

where $a_{0,1}$ are the MHP lattice parameters for $c_{Br} = 0, 1$, respectively.

The local displacement $\mathbf{u}(\mathbf{r})$ in an infinite crystal obeys the equilibrium equation

$$C_{jkmn}\frac{\partial^2 u_m(\mathbf{r})}{\partial x_k \partial x_n} + f_j(\mathbf{r}) = 0, x_{j,k,m,n} = x, y, z; \mathbf{r} = (x, y, z) \tag{A2}$$

where $C_{ijkl}$ are the elastic constants of MHP in the 4-index notation (assumed not dependent on $c_{Br}$), and $\mathbf{f}(\mathbf{r})$ is the volume-force density. In cubic crystals it can be expressed using the gradient of $c_{Br}$:

$$f_j(\mathbf{r}) = -\frac{\partial c_{Br}(\mathbf{r})}{\partial x_j}(C_{11} + 2C_{12})\eta; \tag{A3}$$

here we used the 2-index notation of the elastic constants $C_{11} \equiv C_{xxxx}, C_{12} \equiv C_{xxyy}$. In an infinite crystal, equations (A2, A3) can be solved by Fourier method putting

$$\mathbf{u}^{(FT)}(\mathbf{k}) = \int d^3\mathbf{r}\, \mathbf{u}(\mathbf{r})e^{-i\mathbf{k}\cdot\mathbf{r}}, \tag{A4}$$

which converts the system of partial differential equations (A2) to a set of linear algebraic equations that can be solved directly:

$$\mathbf{u}^{(FT)}(\mathbf{k}) = \widehat{G}(\mathbf{k})\mathbf{f}^{(FT)}(\mathbf{k}), \tag{A5}$$

where $\widehat{G}(\mathbf{k})$ is the elastic Green function defined as

$$\left[\widehat{G}(\mathbf{k})^{-1}\right]_{jm} = C_{jkmn}k_k k_n \tag{A6}$$

In all formulas, the superscript (FT) denotes the Fourier transformation.

**Appendix B. Symmetry of Diffuse Diffracted Intensity from Strain Component**

For the simple case of spherical inclusion or an inclusion with gaussian profile of concentration we can always choose the coordinate system such that the centre of the sphere is the origin of the coordinate system. From the spherical symmetry it is immediately clear that for any strain field $\mathbf{u}(\mathbf{r})$ created by this inclusion $\mathbf{u}(-\mathbf{r}) = -\mathbf{u}(\mathbf{r})$ must be true. If in equation (11) we consider only the strain component we can express the diffuse part of the diffracted intensity with only the strain component from equation (6) as

$$I_h^{(diff,strain)}(\mathbf{q}) = |\langle S_h \rangle|^2 \left| \int d^3r \left[ e^{-i\mathbf{h}\cdot\mathbf{u}(\mathbf{r})} - 1 \right] e^{-i\mathbf{q}\cdot\mathbf{r}} \right|^2 \tag{B1}$$

Therefore, from the symmetry condition $\mathbf{u}(-\mathbf{r}) = -\mathbf{u}(\mathbf{r})$ $I_h^{(diff,strain)}(-\mathbf{q}) = I_h^{(diff,strain)}(\mathbf{q})$ follows, therefore the absolute value of $E_h^{(diff,FT)}(\mathbf{q})$ as well as the diffracted intensity is symmetrical.

Similar symmetry condition can be found for a random distribution of concentration with correlation function given in main text by equation (7) assuming the asymmetry factor $\alpha = 1$. We express the diffuse part of the diffracted intensity with only the strain component as

$$\begin{aligned} I_h^{(diff,strain)}(\mathbf{q}) &= |\langle S_h \rangle|^2 \int\int d^3r\, d^3r'\, e^{-i\mathbf{q}(\mathbf{r}-\mathbf{r}')} \left[ \langle e^{-i\mathbf{h}\cdot(\mathbf{u}(\mathbf{r})-\mathbf{u}(\mathbf{r}'))} \rangle - \langle e^{-i\mathbf{h}\cdot\mathbf{u}(\mathbf{r})} \rangle \right. \\ &\quad \left. - \langle e^{i\mathbf{h}\cdot\mathbf{u}(\mathbf{r}')} \rangle + 1 \right] \end{aligned} \tag{B2}$$

Since $\mathbf{u}(\mathbf{r})$ is a random variable with gaussian distribution and zero mean, we can rewrite the average as $\langle e^{i\mathbf{h}(\mathbf{u}(\mathbf{r})-\mathbf{u}(\mathbf{r}'))} \rangle = e^{-\frac{1}{2}\langle [\mathbf{h}\cdot(\mathbf{u}(\mathbf{r})-\mathbf{u}(\mathbf{r}'))]^2 \rangle}$, and $\langle e^{-i\mathbf{h}\cdot\mathbf{u}(\mathbf{r})} \rangle = \langle e^{i\mathbf{h}\cdot\mathbf{u}(\mathbf{r}')} \rangle$ Substituting this in equation (B2) we find that again $I_h^{(diff,strain)}(-\mathbf{q}) = I_h^{(diff,strain)}(\mathbf{q})$, i.e., the absolute value of $E_h^{(diff,FT)}(\mathbf{q})$ as well as the diffracted intensity is symmetrical even for a random distribution of concentration with $\alpha = 1$. As explained in the main text, the asymmetry of the intensity distribution visible in Figs. 4(a-d) is caused solely by the interference of the strain term $\langle S_h \rangle (e^{-i\mathbf{h}\cdot\mathbf{u}(\mathbf{r})} - 1)$ with the contribution of the chemical contrast $\Delta S_h(\mathbf{r}) e^{-i\mathbf{h}\cdot\mathbf{u}(\mathbf{r})}$.